\documentclass[cameraready]{Interspeech}
\title{Your U-Net Dereverberation Model is Secretly an RIR Encoder}

\author[affiliation={1}]{Sina}{Khanagha}
\author[affiliation={1}]{Timo}{Gerkmann}
\address{
    $^1$ Signal Processing Group, University of Hamburg, Germany 
}

\email{sina.khanagha@uni-hamburg.de}


\usepackage{comment}
\usepackage{amsmath}
\usepackage{graphicx}  
\usepackage{cleveref} % enables \cref
\usepackage{acronym}
\begin{document}
\acrodef{EMA}{exponential moving average}

\maketitle

\begin{abstract}
    In this work, we analyze the ability of NCSN++ U-Net based audio dereverberation models to capture global room characteristics in their intermediate representations. Through an empirical study of both a state-of-the-art diffusion-based model and a discriminative counterpart, we show that deeper layers encode structured room impulse response (RIR)-dependent embeddings. Moreover, the discriminative ability of this implicit room representation correlates with dereverberation performance across objective metrics. Motivated by this observation, we propose a training strategy that explicitly conditions the network on pre-trained RIR embeddings, obtained via self-supervised contrastive learning. Incorporating RIR conditioning improves representation quality, accelerates convergence, and enhances dereverberation performance, while significantly reducing the number of reverse diffusion steps required by the diffusion-based model during inference.

\end{abstract}

\section{Introduction}

The reverberation effect originates from reflection of acoustic waves in closed environments from various surfaces which in turn degrades the intelligibility and the overall perceived quality of speech \cite{reverb_book}. Therefore, the goal of dereverberation models is to revert this effect and retrieve the anechoic speech from the reverberant mixture. More specifically, in this work we will focus on the single-channel case where the audio samples are recorded using a single microphone.

Earlier dereverberation methods relied mainly on statistical assumptions regarding anechoic and reverberant mixture speech to perform the task \cite{ealry_method_habet, early_method_1, early_method_2}. These methods can be further categorized into \textit{informed} \cite{early_method_2} and \textit{blind} \cite{blind_method_1, blind_method_2} depending on the availability of the room impulse response (RIR). Prior knowledge of the RIR in informed scenarios simplifies the task but precise knowledge of the RIR in real-world scenarios is a relatively unrealistic assumption limiting the applicability of informed methods compared to blind counterpart.

Most recent methods utilize deep neural networks (DNNs) and tackle the problem from a data-driven angle \cite{older_dnn_derev, sgmse+, storm, streamfm}. The typical procedure is to train a DNN on pairs of anechoic and reverberant speech pairs in a supervised manner. There are also unsupervised blind methods \cite{buddy, unsup_dereverb_2} that typically show worse performance in matched condition evaluations but demonstrate superior robustness in mismatched conditions (out-of-distribution data). 

Diffusion-based models have gained significant popularity in speech enhancement (SE) tasks in recent years \cite{sgmse+, storm}. The general SE problem is typically formulated as an additive process, which is well aligned with the denoising principles underlying diffusion-based generative models. Nevertheless, it has been empirically demonstrated that diffusion models are also competitive with discriminative baselines in dereverberation tasks and can even exhibit superior robustness under mismatched conditions \cite{buddy}.

A common hypothesis explaining the strong performance of diffusion models in additive SE is that these methods aim to learn the conditional probability paths that progressively transform a noisy sample, drawn from different noise level distributions, into a clean speech sample through multiple refinement steps \cite{sgmse+, storm, streamfm}. However, reverberation arises from a deterministic convolutional process rather than a random additive one. Therefore, the same probabilistic interpretation provided by diffusion-based SE literature does not directly apply. In contrast, discriminative models are typically trained to learn a direct, deterministic mapping between a noisy input and its corresponding clean target which is more aligned with the deterministic convolutional nature of the reverberation theory. Despite the growing success of diffusion-based methods for the task, theoretical or empirical analysis of their probabilistic interpretation under deterministic reverberation has remained largely unexplored.

To better understand the behavior of diffusion-based and discriminative models for dereverberation, we first analyze the internal representations learned by two state-of-the-art models. By extracting features from deeper U-Net layers, we show that both diffusion-based and discriminative approaches learn similar structured, RIR-dependent representations. This observation suggests that diffusion-based baselines in fact exhibit hybrid behavior, combining generative modeling with implicit determinstic RIR encoding. Furthermore, we demonstrate a strong correlation between our diffusion baseline's ability to cluster individual RIRs and its dereverberation performance. Motivated by this finding, we propose a practical strategy to enhance representation learning by conditioning the diffusion-based baseline on RIR embeddings obtained from a pre-trained contrastive RIR encoder. We show that this proof-of-concept approach accelerates convergence, improves representation quality, and significantly reduces the number of reverse diffusion steps needed to reach desirable performance.

\section{Background}
Reverberation in speech is the result of sound reflections from surfaces in an enclosed environment, which results in temporal smearing of the original signal. In signal processing theory, it is commonly modeled as a linear time-invariant convolution between the clean speech $x(t)$ and the RIR $h(t)$:
\begin{equation}
y(t) = x(t) * h(t),
\end{equation}
where $y(t)$ is the observed reverberant signal. The RIR $h(t)$ characterizes the environment's acoustic properties, including reflection delays, decay rates, and frequency-dependent absorption. For simulation, reverberation is applied by convolving clean speech with pre-recorded or synthetic RIRs, enabling controlled evaluation of dereverberation methods.

Even with a known room impulse response (RIR), perfect dereverberation is a non-trivial task since real RIRs are often non-minimum phase, making direct inversion unstable \cite{early_method_2, rir_mixed_phase}. Therefore, DNN-based dereverberation models often aim to directly undo the effect of the convolved RIR and recover the clean speech sample instead of RIR approximation.

Discriminative dereveberation models learn a deterministic mapping 
$f_\theta(\mathbf{y})$ from reverberant speech $\mathbf{y}$ to clean speech 
$\mathbf{x}$ via supervised regression \cite{disc_vs_gen_se}. The training objective is often simply defined as a mean-squared error (MSE):
\begin{align}
\mathcal{L}_{\mathrm{disc}}(\theta)
&= \mathbb{E}_{(\mathbf{x},\mathbf{y})}
\lVert f_\theta(\mathbf{y}) - \mathbf{x} \rVert_2^2 .
\label{eq:disc_loss}
\end{align}

Diffusion-based dereverberation models, in contrast, implicitly learn the conditional distribution $p(\mathbf{x}\mid\mathbf{y})$ of clean speech given a noisy observation. During training, the model predicts the Gaussian noise added at each diffusion step \cite{sgmse+}. By learning these conditional denoising steps, diffusion models approximate the full distribution of clean speech and enable iterative reconstruction.
% %
% \begin{align}
% \mathcal{L}_{\mathrm{diff}}(\theta)
% &= \mathbb{E}_{\mathbf{x},\boldsymbol{\epsilon}\sim\mathcal{N}(\mathbf{0},\mathbf{I}),t}
% \lVert 
% \boldsymbol{\epsilon}
% -
% \boldsymbol{\epsilon}_\theta(\mathbf{x}_t,\mathbf{y},t)
% \rVert_2^2 .
% \label{eq:diff_loss}
% \end{align}
% %

However, an important design choice in many of the recent diffusion-based dereverberation models is that the reverberant speech $\mathbf{y}$ itself is also concatenated with the current process state $\mathbf{x}_t$ as one of the input channels to the U-Net DNN at every time step $t$ \cite{sgmse+, storm, streamfm}. We argue that availability of the original reverberant sample at every step of diffusion process inherently results in a hybrid model that combines discriminative deterministic mapping functions with the conditional generative properties of diffusion models. We will provide a more in-depth discussion about this theory in \cref{sec:results}.

\begin{figure*}[t]
\centering
\includegraphics[width=\textwidth]{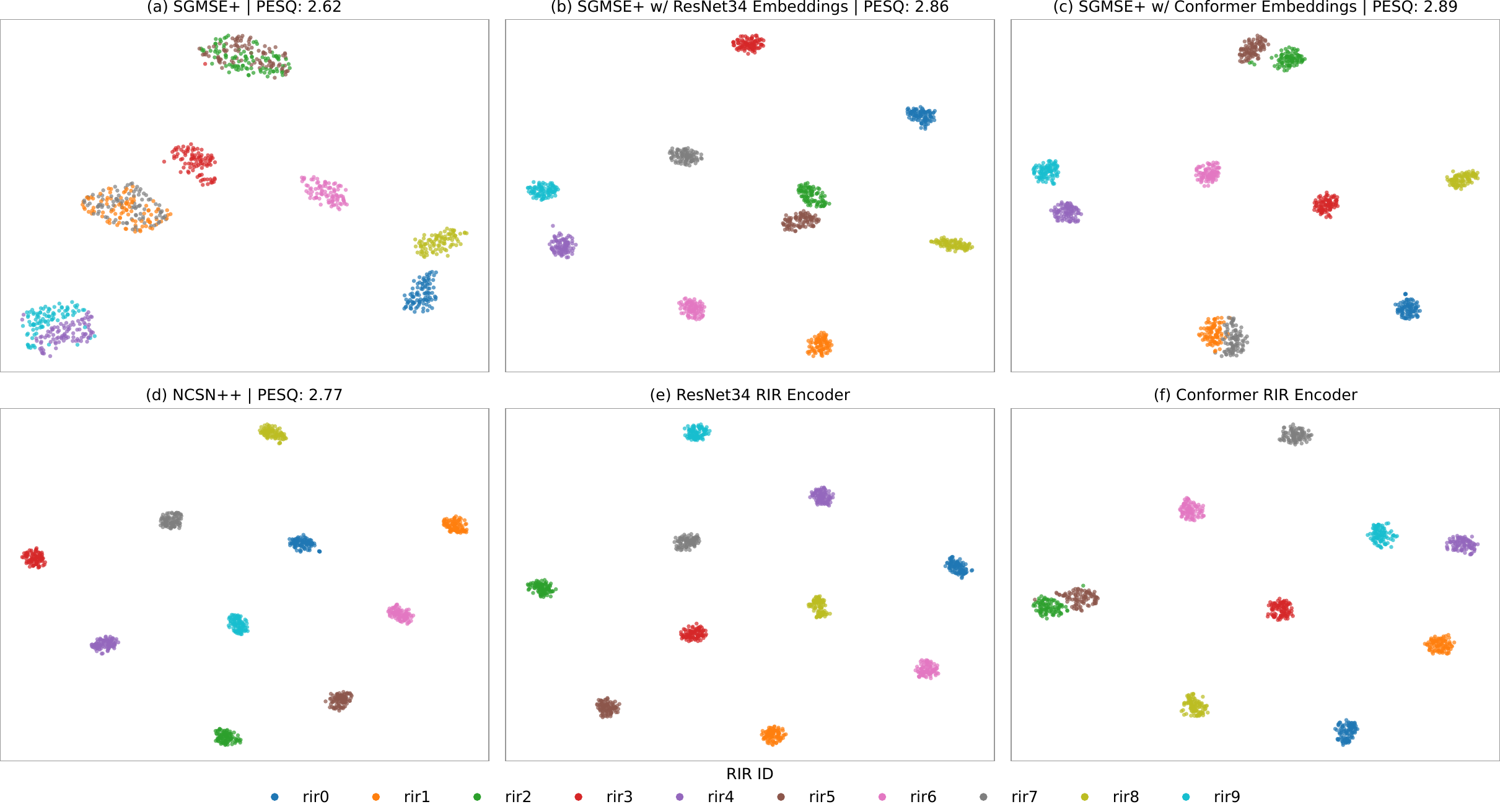}
\caption{t-SNE visualizations of embeddings extracted from 950 simulated reverberant samples. For the dereverberation models (a–d), embeddings are taken from the attention block of the NCSN++ backbone. For the proposed RIR encoders (e–f), embeddings are obtained directly from the encoder outputs. The corresponding dereverberation performance on the VCTK-Reverb test set (Figure~\ref{fig:metrics-fig}), with N=50 for diffusion-based models, is also reported for each model.}
\label{fig:tsne-fig}
\end{figure*}

\section{Method}
\subsection{RIR Encoder Training}

Inspired by recent work in this field \cite{cl_param_est, cl_rir_est, cl_multichannel, cl_av_rir, rev_rir}, we train the RIR encoder using a contrastive learning framework that encourages embeddings from the speech samples corrupted by the same RIR to be close while pushing embeddings from different RIRs apart. For each training batch, two speech utterances are convolved with the same RIR to produce a positive pair. The encoder outputs fixed-length embeddings for each convolved signal, which are then normalized and compared using a cosine similarity-based InfoNCE loss \cite{infoNCE}:

\begin{align}
\mathcal{L}_{\mathrm{pos}}
&=
- \frac{1}{B}
\sum_{i=1}^{B}
\log \left(
\frac{
\sum_{j \in \mathcal{P}(i)}
\exp\left(
\frac{\mathbf{z}_i^\top \mathbf{z}_j}{\tau}
\right)
}{
\sum_{\ell \neq i}
\exp\left(
\frac{\mathbf{z}_i^\top \mathbf{z}_\ell}{\tau}
\right)
}
\right).
\end{align}
where $B$ is the batch size, $\mathbf{z}_i \in \mathbb{R}^d$ denotes the $\ell_2$-normalized embedding of sample $i$, $\mathcal{P}(i)$ denotes the set of positive indices for anchor $i$, and $\tau$ is a temperature hyperparameter. A secondary hard negative-pair loss is also included to explicitly encourage the model to ignore the speech content of a sample. Hard negatives are created by convolving the same speech utterance with two different RIRs and trying to minimize the similarity between the output embeddings of these two samples:
\begin{align}
\mathcal{L}_{\mathrm{neg}}
&=
\frac{1}{B}
\sum_{i=1}^{B}
\mathbf{z}_i^\top \mathbf{z}_i^{-}.
\end{align}
Since the embeddings in both loss terms are $\ell_2$-normalized, the loss reduces to cosine similarity in the range of
\(
[-1,1]
\). The total training objective is a weighted combination of these terms:
\begin{align}
\mathcal{L}_{\text{total}}
&=
\mathcal{L}_{\text{pos}}
+
\lambda \mathcal{L}_{\text{neg}} .
\label{eq:total_loss}
\end{align}
enabling the encoder to learn a rich embedding space that captures the variability of RIRs while remaining invariant to the convolved speech content.

\subsection{FiLM Conditioning}

Both discriminative and diffusion-based models commonly employ U-Net based architectures. In particular, the NCSN++ U-Net, originally proposed by \cite{song_score} for diffusion-based image generation, has demonstrated strong performance in both SE and dereverberation tasks. As a result, it has become a popular backbone for state-of-the-art SE and dereverberation models. Both of our baseline models adopt the NCSN++ architecture with the exact configurations specified by the original authors, resulting in negligible differences in model size across the baselines.

Feature-wise Linear Modulation (FiLM) \cite{film} is a conditioning mechanism that adapts intermediate network features using external information. Given an intermediate feature map $\mathbf{F}\in \mathbb{R}^{C \times T}$ and a conditioning vector $c$, a small conditioning network learns channel-wise scaling and shifting parameters:

\begin{align}
\boldsymbol{\gamma} &= f_{\gamma}(\mathbf{c}), 
\qquad
\boldsymbol{\beta} = f_{\beta}(\mathbf{c}),
\label{eq:film_params}
\end{align}
where $\boldsymbol{\gamma}, \boldsymbol{\beta} \in \mathbb{R}^{C}$. 
The modulated features are then given by
\begin{align}
\mathrm{FiLM}(\mathbf{F} \mid \mathbf{c})
&= \boldsymbol{\gamma} \odot \mathbf{F}
+ \boldsymbol{\beta}.
\label{eq:film_mod}
\end{align}

We inject the RIR embedding into each BigGAN-style \cite{biggan} residual block of NCSN++ at every resolution using a FiLM-based modulation applied after the second normalization layer. The conditioning embedding is first normalized via LayerNorm and passed through a linear projection to produce channel-wise scaling and shifting parameters, which are split into $\gamma$ and $\beta$. To ensure stable training, the FiLM layer is initialized to behave as an identity mapping by zero-initializing its weights and biases. Furthermore, the modulation parameters are bounded by $\gamma = 1 + 0.1 \tanh(\gamma_{\text{raw}})$ and $\beta = 0.1\,\beta_{\text{raw}}$, limiting the magnitude of the initial perturbation.

\section{Experimental Setup}
\subsection{Dataset}
We use the VCTK corpus \cite{vctk} as clean speech, selecting 103 speakers for training, two for validation, and two for testing. The dataset contains approximately 44 hours of audio, which we downsample to 16 kHz for all experiments. Reverberant conditions are simulated using around 10K real RIRs collected from multiple public datasets \cite{buddy}, which we split into training, validation, and test sets with a 0.9/0.05/0.05 ratio. We refer to this dataset as VCTK-Reverb.

Naturally, due to the fact that in our VCTK-Reverb test set each RIR has been used 1 to 3 times, it is difficult to directly conduct RIR representation analysis on the data. Therefore, in order to investigate and visualize the RIR embeddings of our RIR encoders and attention features of dereverberation baselines, we created a secondary dataset. For this dataset, we randomly select 95 clean utterances from the two VCTK test speakers and 10 RIRs from the test set and convolve each RIR with every utterance, resulting in 950 reverberant samples. 

All reported speech quality metrics are computed strictly on the 774 samples of the VCTK-Reverb test set; the secondary dataset is used only for embedding visualization.

\subsection{Baselines}
For our experiments, we selected two dereverberation models: the diffusion-based SGMSE+ \cite{sgmse+} and NCSN++ U-Net as a discriminative baseline, which also serves as the backbone for the diffusion model. We train SGMSE+ using the official implementation provided by the authors and adopt all default training hyperparameters as described in the original work \cite{sgmse+}, unless stated otherwise. Note that we use the smaller SGMSE+M and NCSN++M configurations \cite{storm} with 27.8M parameters. For simplicity, we omit the suffix “M” throughout the paper.

Furthermore, we used wide-band PESQ \cite{pesq} and DNSMOS \cite{dnsmos} as comparison metrics. All models are trained from scratch until convergence, and for evaluation, we selected the checkpoint achieving the highest wide-band PESQ on a smaller subset of validation samples.
%TODO: BATCHSIZE AND OTHER STFF%

Additionally, we train two RIR encoders: a ResNet34-based encoder \cite{resnet} and a Conformer-based encoder \cite{conformer} (10 layers, 256 hidden units, 4 attention heads), both followed by a two-layer projection head producing 256-dimensional $\ell_2$-normalized embeddings. Both models were trained using the same modified InfoNCE objective ($\lambda = 0.2$ and $\tau = 0.07$) with AdamW (learning rate $10^{-4}$, weight decay $10^{-2}$, batch size 16) for 200 epochs, with the best model selected based on validation loss. Complete architectural specifications and training hyperparameters are available in the accompanying repository\footnote{\url{https://github.com/sp-uhh/rir-encoder}}.

\section{Results and Discussion}\label{sec:results}
\subsection{RIR Representation Analysis}
For this analysis, reverberant speech samples are passed through the model, and embeddings are extracted from the attention block of NCSN++ via global average pooling over the spatial dimensions.
Let $\mathbf{F}_{\text{att}} \in \mathbb{R}^{C \times H \times W}$ denote the attention feature tensor, where $C$ is the number of channels and $H$ and $W$ are the spatial dimensions. The corresponding embedding is obtained via global average pooling:
\begin{align}
\mathbf{e}_{\text{att}}
&= \frac{1}{HW}
\sum_{h=1}^{H} \sum_{w=1}^{W}
\mathbf{F}_{\text{att}}(:,h,w) ,
\label{eq:att_embedding}
\end{align}
where $\mathbf{e}_{\text{att}} \in \mathbb{R}^{C}$.
Subsequently, we apply the unsupervised t-SNE \cite{tsne} clustering algorithm on the extracted embeddings. The results are illustrated in Figure~\ref{fig:tsne-fig}. From the first column of the figure, we can clearly observe that, regardless of the training paradigm or loss function, both of our baseline methods learn a strong global representation of the degradation operator (RIR). The resulting clusters closely resemble an RIR encoder. 

Notably, these clusters begin to emerge in the early layers, become increasingly stronger toward the bottleneck and attention blocks of NCSN++, and gradually diminish near the final layers. However, this behavior is expected since the initial and output representations are closer to the raw STFT domains of the input and reconstructed speech, where features are more strongly influenced by speech content than by the underlying RIR. We chose to visualize the attention block features as we observed the strongest clusters at these blocks. 

Moreover, the plots suggest that diffusion models exhibit a hybrid behavior. Specifically, they appear to learn deterministic representations associated with individual RIRs, largely independent of the diffusion timestep $\mathbf{t}$ and process state $\mathbf{x_t}$. We argue that since our diffusion baseline explicitly provides the reverberant sample as one of the input channels at every time step, the learned representations exhibit deterministic, degradation-specific structure, while the sampling process preserves generative characteristics. At the same time, they generate perceptually more natural outputs, likely due to the stochastic generative mechanism of the diffusion process. We hypothesize that this combination of structured degradation encoding and stochastic refinement contributes to the adaptability of diffusion models to the non-additive nature of reverberation problem.

The first row also indicates that there may be a correlation between the strength of the model in creating well-delimited clusters and its ability in dereverberation task. We can observe that the SGMSE+ trained with the \ac{EMA} of 0.99 has simultaneously the weakest clustering ability and weakest dereverberation performance of the three models. In the following section we will further investigate this theory by trying to enhance the RIR representation learning of the SGMSE+ model.

% \begin{table}[t]
% \centering
% \caption{VCTK-Reverb test set dereverberation performance comparison.}
% \label{tab:main_results}
% \resizebox{\columnwidth}{!}{
% \begin{tabular}{l|c c c}
% \textbf{Model} & \textbf{PESQ} $\uparrow$ & \textbf{ESTOI} $\uparrow$ & \textbf{DNSMOS} $\uparrow$ \\
% \hline
% Mixture & - & - & - \\
% FM & 3.38 ± 0.43 & 0.86 ± 0.08 & 3.66 ± 0.23 \\
% NCSN++ & 2.77 ± 0.51 & 0.79 ± 0.10 & 3.51 ± 0.26 \\
% \hline
% SGMSE+ & 2.43 ± 0.42 & 0.71 ± 0.10 & 3.54 ± 0.20 \\
% % \quad + EMA & & & \\
% SGMSE+ w/ Conformer & \textbf{2.68} ± 0.45 & \textbf{0.73 ± 0.10} & \textbf{3.62 ± 0.23} \\
% % \quad + EMA & & & \\
% SGMSE+ w/ ResNet34  & 2.58 ± 0.43 & 0.72 ± 0.10 & 3.59 ± 0.23 \\
% \hline
% \end{tabular}
% }
% \end{table}

% \begin{table}[t]
% \centering
% \caption{Effect of EMA on dereverberation performance of SGMSE+ on VCTK-Reverb test set}
% \label{tab:EMA_results}
% \resizebox{\columnwidth}{!}{
% \begin{tabular}{l|c c c}
% \textbf{Model} & \textbf{PESQ} $\uparrow$ & \textbf{ESTOI} $\uparrow$ & \textbf{DNSMOS} $\uparrow$ \\
% \hline
% SGMSE+ & 2.43 ± 0.42 & 0.71 ± 0.10 & 3.54 ± 0.20 \\
% \quad - EMA & 2.24 ± 0.36 & 0.69 ± 0.10 & 3.48 ± 0.24 \\
% \hline
% SGMSE+ w/ Conformer & 2.40 ± 0.39 & 0.70 ± 0.10 & 3.54 ± 0.24 \\
% \quad - EMA & \textbf{2.68} ± 0.45 & \textbf{0.73 ± 0.10} & \textbf{3.62 ± 0.23}\\
% \hline
% \end{tabular}
% }
% \end{table}

\begin{figure}[t]
\centering
\includegraphics[width=\columnwidth]{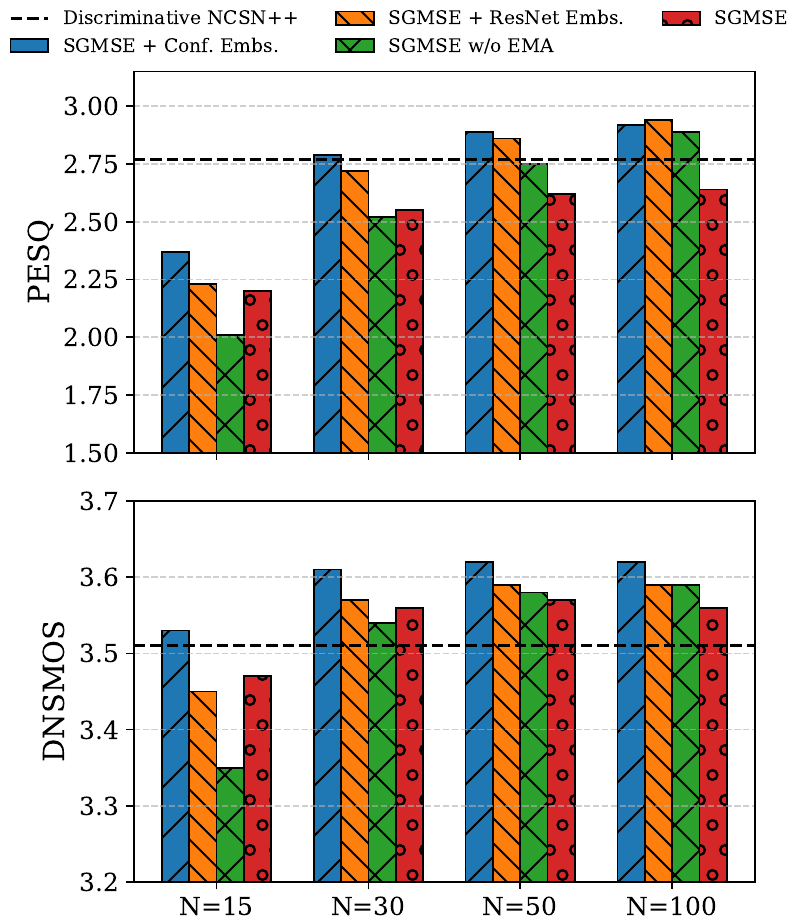}
\caption{PESQ and DNSMOS for different models across various reverse diffusion steps (N).}
\label{fig:metrics-fig}
\end{figure}

\begin{figure}[t]
\centering
\includegraphics[width=\columnwidth]{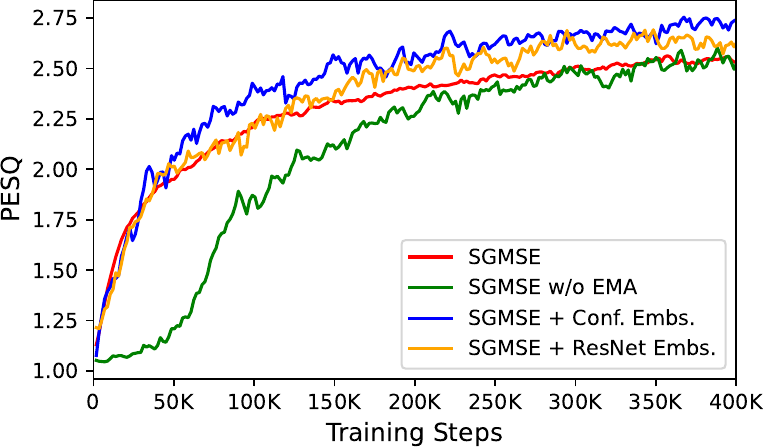}
\caption{Validation PESQ over training steps.}
\label{fig:train-fig}
\end{figure}

\subsection{Improving Representation Learning}
In this section, we will investigate whether conditioning the backbone U-Net on pre-trained RIR embeddings can improve the dereverberation performance of the model. First, (e) and (f) plots from Figure~\ref{fig:tsne-fig} verify the fact that our pre-trained RIR encoders are able to produce similar contextual representations to representations learned by the dereverberation models.

Comparing Figure~\ref{fig:tsne-fig}(a) with (b) and (c), conditioning SGMSE+ on pre-trained RIR embeddings leads to more structured and separable RIR-dependent clusters in the latent space. As shown in Figure~\ref{fig:metrics-fig}, this representation improvement is accompanied by consistent PESQ gains of 0.17, 0.24, 0.27, and 0.28 for different numbers of reverse diffusion inference steps N for the Conformer-conditioned model. A similar trend is also observable for DNSMOS metric for both of our RIR-conditioned models. These results support our hypothesis that strengthening RIR discriminability is associated with improved dereverberation performance.

We also want to highlight an important detail in training SGMSE+ with our proposed RIR embeddings which is the fact that the \ac{EMA} should be excluded from training for results to improve. Conversely, setting \ac{EMA} to zero for the baseline SGMSE+ has detrimental effects on its convergence (Figure~\ref{fig:train-fig}) and the number of required steps to reach desirable performance (Figure~\ref{fig:metrics-fig}). However, with much longer training and twice as many reverse diffusion steps at inference compared to our RIR-conditioned models, it does outperform the baseline SGMSE and demonstrate competitive performance with our proposed models. The inferior performance of SGMSE+ baseline compared to our models could also be potentially explained by the dependence of SGMSE+ baseline on \ac{EMA} for stable training limiting its learning capabilities. Evidently, our conditioning method solves this optimization issue and stabilizes the optimization in earlier stages (Figure~\ref{fig:train-fig}). A summary of the results is presented in Figure~\ref{fig:metrics-fig}. 

\section{Conclusion}

In this work, we investigated the RIR representation learning capability of diffusion-based and discriminative NCSN++ U-Net–based dereverberation models by applying unsupervised clustering to spatially averaged embeddings extracted from mid-level attention features. Our analysis reveals that these models implicitly learn strong RIR-dependent representations. Additionally, we observed a correlation between the discriminative strength of learned representations and dereverberation performance. Motivated by this observation, we proposed a conditioning strategy in which the network is trained with pre-trained RIR embeddings learned via a contrastive framework. This approach leads to improved RIR representation learning, faster convergence, and enhanced dereverberation performance.

\section{Acknowledgments}
Funded by the Deutsche Forschungsgemeinschaft (DFG, German Research Foundation) – 545210893, 498394658. The authors gratefully acknowledge the scientific support and HPC resources provided by the Erlangen National High Performance Computing Center (NHR@FAU) of the Friedrich-Alexander-Universitat Erlangen-Nurnberg (FAU) under the NHR project f102ac. NHR funding is provided by federal and Bavarian state authorities. NHR@FAU hardware is partially funded by the German Research Foundation (DFG) – 440719683. 
\section{Generative AI Use Disclosure}
Generative AI tools were used only for minor language editing (clarity, grammar, and polishing). All ideas, methods, experiments, results, and conclusions were produced and verified by the authors, who take full responsibility for the paper. 

\bibliographystyle{IEEEtran}
\bibliography{mybib}

\end{document}